\documentclass[sigconf]{acmart}
\AtBeginDocument{%
  \providecommand\BibTeX{{%
    \normalfont B\kern-0.5em{\scshape i\kern-0.25em b}\kern-0.8em\TeX}}}

\usepackage{balance}
\usepackage{multirow}
\usepackage{subfigure}
\usepackage{multirow}
\usepackage{subfigure}
\usepackage{geometry}
\newcommand{\jiayan}[1]{MSGIFSR}

\copyrightyear{2022} 
\acmYear{2022} 
\setcopyright{acmcopyright}\acmConference[WSDM '22]{Proceedings of the Fifteenth ACM International Conference on Web Search and Data Mining}{February 21--25, 2022}{Tempe, AZ, USA}
\acmBooktitle{Proceedings of the Fifteenth ACM International Conference on Web Search and Data Mining (WSDM '22), February 21--25, 2022, Tempe, AZ, USA}
\acmPrice{15.00}
\acmDOI{10.1145/3488560.3498524}
\acmISBN{978-1-4503-9132-0/22/02}

\begin{document}
\title{Learning Multi-granularity Consecutive User Intent Unit \\ for Session-based Recommendation}


\author{Jiayan Guo$^{1*}$, Yaming Yang$^{2}$, Xiangchen Song$^{3*}$, Yuan Zhang$^{1}$, Yujing Wang$^{2}$\\ Jing Bai$^{4}$, Yan Zhang$^{1}$}\thanks{${^*}$This work was done when Jiayan Guo and Xiangchen Song were interning at Microsoft Research Lab Asia, Beijing, China}
\affiliation{
$^{1}$School of Artificial Intelligence, Peking University\country{China}, $^{2}$Microsoft Research Asia\country{China}, \\ $^{3}$Carnegie Mellon University\country{USA}, $^{4}$Microsoft\country{China}
}

\email{{guojiayan,yuan.z,zhyzhy001}@pku.edu.cn}
\email{{yayaming, yujwang, jbai}@microsoft.com}
\email{{xiangchensong}@cmu.edu}

\renewcommand{\shortauthors}{Jiayan Guo and Yaming Yang, et al.}


\begin{abstract}

Session-based recommendation aims to predict a user's next action based on previous actions in the current session. The major challenge is to capture authentic and complete user preferences in the entire session. Recent work utilizes graph structure to represent the entire session and adopts Graph Neural Network (GNN) to encode session information. This modeling choice has been proved to be effective and achieved remarkable results. However, most of the existing studies only consider each item within the session independently and do not capture session semantics from a high-level perspective. Such limitation often leads to severe information loss and increases the difficulty of capturing long-range dependencies within a session.

Intuitively, compared with individual items, a session snippet, i.e., a group of locally consecutive items, is able to provide supplemental user intents which are hardly captured by existing methods. In this work, we propose to learn multi-granularity consecutive user intent unit to improve the recommendation performance. Specifically, we creatively propose \textit{Multi-granularity Intent Heterogeneous Session Graph}~(MIHSG) which captures the interactions between different granularity intent units and relieves the burden of long-dependency. Moreover, we propose the \textit{Intent Fusion Ranking}~(IFR) module to compose the recommendation results from various granularity user intents. Compared with current methods that only leverage intents from individual items, IFR benefits from different granularity user intents to generate more accurate and comprehensive session representation, thus eventually boosting recommendation performance. We conduct extensive experiments on five session-based recommendation datasets and the results demonstrate the effectiveness of our method. Compared to current state-of-the-art methods, we achieve as large as 10.21\% gain on HR@20 and 15.53\% gain on MRR@20. Code is available at~\url{https://github.com/SpaceLearner/SessionRec-pytorch}.



\end{abstract}





\maketitle



\section{Introduction}

With the rapid growth of the amount of information on the Internet, massive products, contents, and services (which are uniformly described as items) are emerging every day. It becomes difficult for users to view all items due to the time limit. Thus \textit{Recommender Systems} (RS) have played an important role in helping make efficient and satisfying choices.
As is pointed out, user preference tends to be dynamic, hence short-term user history captures more accurate user intent~\cite{DBLP:journals/corr/abs-1902-04864}. In such scenario, session-based recommendation~\cite{song2019session,5954341,10.1145/3331184.3331210,10.1145/3397271.3401273,10.1145/3109859.3109872}, which encapsulates a range of latest consecutive user-item interactions as sessions, draws increasing attention and leads to better performance

Traditional session-based methods treat each session as a sequence of items sorted by click time and widely adopt \textit{Recurrent Neural Networks} (RNNs)~\cite{DBLP:journals/corr/HidasiKBT15,DBLP:conf/recsys/TanXL16,DBLP:conf/cikm/LiRCRLM17} to solve the recommendation problem.
Although remarkable performance has been achieved, these methods are arguably insufficient to obtain accurate user representations in sessions and neglect complex transitions of items~\cite{DBLP:conf/aaai/WuT0WXT19}. Instead, recent work~\cite{DBLP:conf/aaai/WuT0WXT19,DBLP:journals/corr/abs-1909-04276,DBLP:conf/ijcai/XuZLSXZFZ19,DBLP:conf/cikm/PanCCCR20,DBLP:conf/kdd/ChenW20,wang2021session} utilizes graph structure to represent the session and employ \textit{Graph Neural Networks} (GNNs) \cite{Kipf:2016tc, DBLP:journals/corr/abs-1812-08434} to conduct information propagation between adjacent items. Results on various academic session-based recommendation datasets demonstrate the significant superiority of GNN based methods over traditional methods.

\begin{figure}[h]
\setlength{\belowcaptionskip}{-1cm}
 \centering
 \includegraphics[width=.7\linewidth]{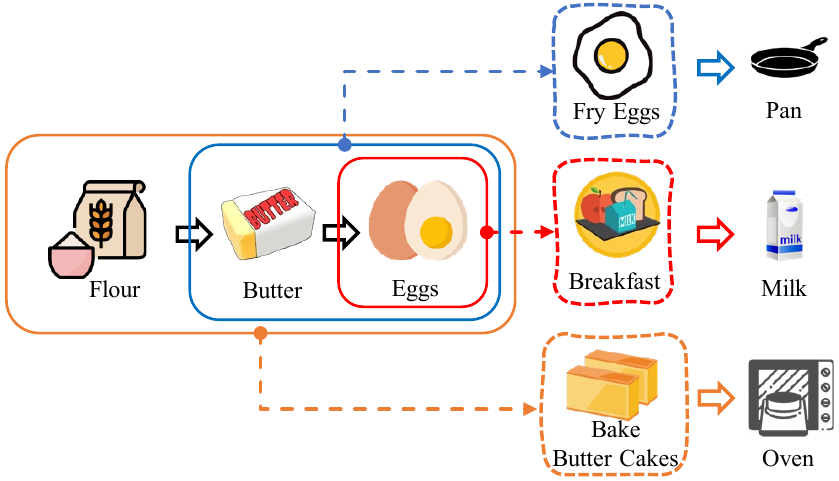}
  \caption{An example of hidden user intents revealed by groups of consecutive items. A session (``Flour'', ``Butter'', ``Eggs'') is given, ``Fry Eggs'', ``Breakfast'' and ``Bake Butter Cakes'' are user intents revealed by grouping different consecutive items in the session. Items in the right are possible items the user will click next under each intent.}
  \label{fig:products}
\end{figure}

However, most of these GNN based methods only take individual items as basic units to extract user preference and ignore the rich hidden user intents revealed by a group of consecutively adjacent items (which is modeled as intent units). Such intent units encapsulate local context-aware user preferences, thereby enabling the recommendation system to generate comprehensive and accurate recommendation results in multiple granularities. For example, in Figure \ref{fig:products}, given a session with (``Flour'', ``Butter'', and ``Eggs''), when focusing on the last item, eggs, the user may want to buy some foods for breakfast. In this case, a reasonable recommendation for his or her next purchase is milk. At the same time, if the intent unit with length two is considered, the local context with ``butter'' and ``egg'' reflects that the user wants to fry eggs. In this scenario, a new pan for cooking is a better suggestion for his or her next purchase. Furthermore, if all three items (``Flour'', ``Butter'', and ``Eggs'') are considered, the intent for such user is obviously changed as the user wants to make a cake or bread, hence an oven is a more suitable recommendation. The example described above shows that a group of consecutive items enriches user intents, and different group granularities reveal different user intents and can help to provide multiple recommendation candidates.. However, 1) \textit{how to model the interaction among these grouped intents} and 2) \textit{how to ensemble these user intents from different granularities} are not trivial.

To tackle these two challenges, we first introduce a novel intent extraction method, i.e., mining user intents both from individual items and combined locally consecutive items. Specifically, besides modeling items independently, we also compose consecutive item groups with different lengths as consecutive intent units~(CIUs) and then model the transition relationships among these CIUs by a \textit{Multi-granularity Intent Heterogeneous Session Graph}~(MIHSG). 
In this heterogeneous graph~\cite{xie2021learning}, nodes~(CIUs) with different numbers of items are categorized into different groups and the transition edges among the same type of nodes capture the spatial continuity of the user-item interactions in the corresponding intent granularity. We also introduce a special type of edge to represent the transition between high order intent units and the single items, which explicitly encode the intent evolution between coarse and fine-grained granularities. 
Compared with existing session graphs, the proposed heterogeneous graph can extract richer user preference, which contributes to alleviate the information loss during the graph modeling process. We then apply Heterogeneous Graph Attention network~(HGAT) to extract the node representations~\cite{ragesh2021hetegcn,velickovic2018graph,wang2019heterogeneous,yang2020heterogeneous,zhang2019heterogeneous,lv2021we}. Since the high-level intents can provide compressed yet accurate session intent information, such information can be propagated more effectively through different intent units. The long-range dependency problem can thus be alleviated. To solve the second problem, we propose \textit{Intent Fusion Ranking}~(IFR), i.e., utilizing last attention to learn session representation for each intent granularity level and compose recommendation results to predict the next item. We name our model  MSGIFSR~(\underline{M}ulti-granularity Intent Heterogeneous \underline{S}ession \underline{G}raph and \underline{I}ntent \underline{F}usion Ranking for \underline{S}ession-based \underline{R}ecommendation). 
Empirically, our method significantly improves the model's performance by enriching contextual user intent within one session and achieves state-of-the-art performance on five benchmark datasets.



In summary, our contributions in this paper are as follows:
\begin{itemize}
    \item We propose consecutive intent unit (CIU) to extract user intent from different granularities. Newly formed consecutive intent units contain more accurate user preference and provide supplementary information for recommendation.
    \item To better exploit CIU, we propose \textit{Multi-granularity Intent Heterogeneous Session Graph} (MIHSG) to model complex transition relationship between different granular intent units explicitly. The MIHSG can propagate information efficiently, especially in long sessions. We also propose \textit{Intent Fusion Ranking}~(IFR) strategy to fully utilize the intent representation from all granularity levels to enhance the recommendation performance.
    \item We compare the performance of MSGIFSR against state-of-the-art baselines on five public benchmark datasets. The results show the superiority of MSGIFSR over the state-of-the-art models, i.e., as large as 10.21\% gain on HR@20 and 15.53\% gain on MRR@20.

\end{itemize}


\begin{figure*}[h]
\setlength{\abovecaptionskip}{0.2cm}
\setlength{\belowcaptionskip}{-0.45cm}
  \centering
  \includegraphics[width=.95\textwidth]{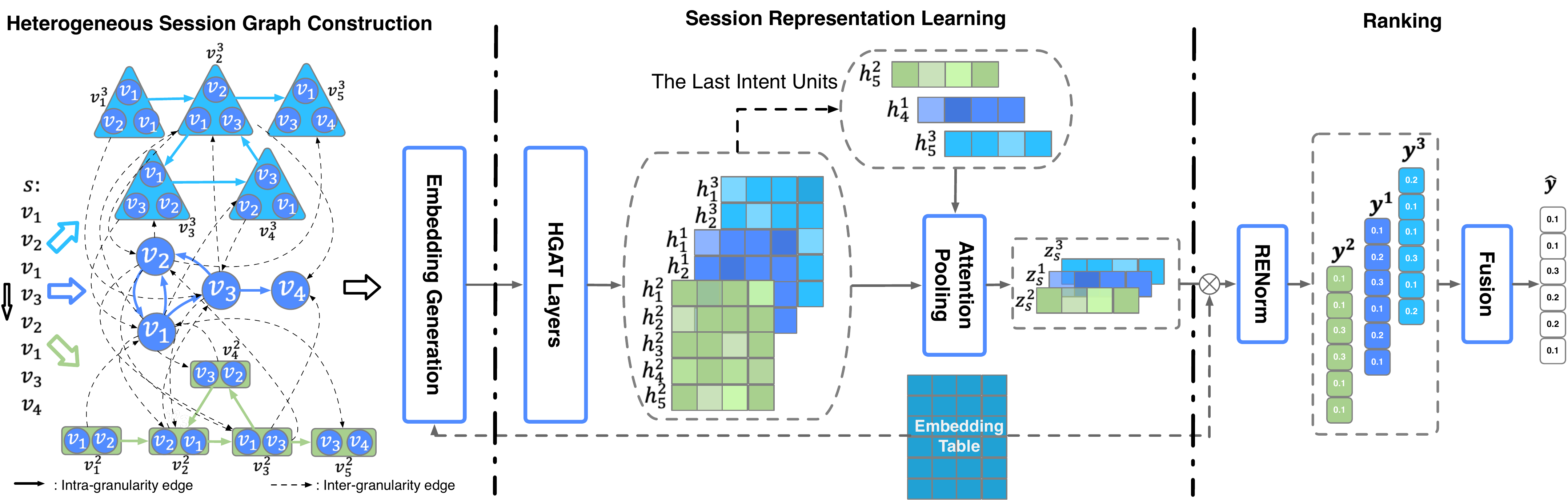}
  \caption{Overview of the MSGIFSR framework. Given a session, we first construct the multi-granularity intent unit  heterogeneous session graph (MIHSG). Then the MIHSG is passed to a heterogeneous graph attention network to get the representations for intent unit from all granularity levels. After that, the intent fusion ranking ~(IFR) module combines session representations from all intent unit representations granularity levels to get the recommendation results.} 
  \label{fig:framework}
\end{figure*}
\section{Related Work}

In this section, we review the related work for session-based recommendation.

Inspired by the fact that similar users tend to buy similar items, the earliest session-based methods are mostly based on nearest neighbors~\cite{10.1145/1864708.1864770, 10.1109/ICTAI.2013.120,5954341}. These methods need a similarity function to compute similarity scores. Then some approaches notice that exploiting sequential behavior is beneficial to predict the next item. Some of them employ Markov chain ~(MC) to capture the sequential signal of interactions. For example, Rendle et al.~\cite{10.1145/1772690.1772773} propose factorization of personalized Markov chains ~(FPMC) to capture both sequential user behavior and long-term interest. King et al.~\cite{10.1145/2766462.2767694} propose a hierarchical representation model (HRM) to improve FPMC with a hierarchical architecture. However, Markov chain-based methods are usually unable to capture more complex higher-order sequential patterns.

The rapid growth of deep learning methods brings significant performance gain on session-based recommendation. Recently proposed neural-based methods employ sequential structure neural networks like RNNs~\cite{hochreiter1997long,cho-etal-2014-learning} to capture the user's sequential behavior. For instance, Bal{\'{a}}zs et al.~\cite{DBLP:journals/corr/HidasiKBT15} propose GRU4Rec to use the gated recurrent unit (GRU)~\cite{cho-etal-2014-learning} to model user sequential behavior, Tan et al.~\cite{DBLP:conf/recsys/TanXL16} apply data augmentation and generalized distillation to improve GRU4Rec and Li et al.~\cite{DBLP:conf/cikm/LiRCRLM17} propose to use attention mechanism to improve the representation capacity of RNNs. In addition, as sessions can be very limited as in-progress information, the neighbor information is introduced to assist in modeling ongoing sessions~\cite{10.1145/3109859.3109872,10.1145/3397271.3401273,10.1145/3331184.3331210}. For example, Jannach and Ludewig~\cite{10.1145/3109859.3109872} introduce a neighborhood session using the K-nearest neighbor (KNN) method. Unlike traditional KNNs, Wang et al.~\cite{10.1145/3331184.3331210} propose CSRM to incorporate neighbor sessions information via memory network~\cite{NIPS2015_8fb21ee7, weston2014memory}. To learn better session representation, in NARM~\cite{DBLP:conf/cikm/LiRCRLM17} an attention pooling method is proposed and is widely applied in subsequent neural recommendation methods~\cite{DBLP:conf/kdd/LiuZMZ18,DBLP:conf/cikm/PanCCCR20,DBLP:conf/ijcai/XuZLSXZFZ19,wang2021session, DBLP:conf/kdd/ChenW20,DBLP:conf/sigir/Yu0LWWT20}. Specifically, these methods use global and local preference to make a recommendation. However, they still consider single items as the base recommendation units thus lack the description of in session combined information. 


Graph neural network (GNN) is introduced to capture the complex transition relationships of sessions~\cite{wang2021session, DBLP:conf/ijcai/XuZLSXZFZ19,DBLP:conf/kdd/ChenW20,DBLP:conf/aaai/WuT0WXT19,DBLP:conf/cikm/PanCCCR20}. For instance, SR-GNN~\cite{DBLP:conf/aaai/WuT0WXT19} takes transition relations between adjacent items to construct unweighted session graphs and uses gated graph neural network ~(GGNN)~\cite{li2016gated} to extract the transition information. Based on this work, Xu et al.~\cite{DBLP:conf/ijcai/XuZLSXZFZ19} propose a method that uses a GGNN to extract local context information and a self-attention network ~(SAN) to capture global dependencies between distant positions. These GNN-based methods have shown a new and promising direction for session-based recommendation, while the constructed session graph faces a lossy session encoding problem. To alleviate such problem, Chen and Wong~\cite{DBLP:conf/kdd/ChenW20} propose lossless edge-order preserving aggregation and shortcut graph attention to efficiently aggregate information and make a recommendation. To propagate information more efficiently in long sessions, Pan et al.~\cite{DBLP:conf/cikm/PanCCCR20} propose to use a star node to bridge items. The newly added star node can filter out irrelevant items, thus make better recommendations on long sessions. However, all these works take single items as intent unit and ignore the consecutive combined user behavior of a group of items, thus losing the recommendation accuracy.  

Some works employ \textit{Hypergraphs}~\cite{Feng_You_Zhang_Ji_Gao_2019} to enhance item representation~\cite{xia2021hyper, wang2021session}. Xia et al.~\cite{xia2021hyper} take each session as a hyperedge to model cross-session item relationship, and Wang et al.~\cite{wang2021session} propose SHARE that uses a sliding window to construct a hypergraph of a single session to capture group intent. However, such hypergraph structure loses the user sequential behavior information, i.e., the sequential order of items. In addition, the last attention pooling only uses the item representation to generate session representation which can not fully utilize the user intent information from all intent unit granularity levels.

\section{Methodology}

In this section, we first introduce the formal definition of the general session-based recommendation problem ~(Section 3.1). Then we describe the proposed multi-granularity consecutive intent unit ~(Section 3.1). Afterwards, we elaborate on three components of the proposed model i.e. the Multi-granularity Intent Heterogeneous Session Graph ~(Section 3.3), the Heterogeneous Graph Attention Network ~(Section 3.4), and the Intent Fusion Ranking ~(Section 3.5). 

\subsection{Problem Definition}

Session-based recommendation is to predict the next-item based on a sequence of clicked items. Formally speaking, let $I=\{v_1, v_2,...v_{|I|}\}$ represent all items in the dataset. A session $s_i=\{v_{{t_1}},v_{{t_2}},...v_{{t_L}}\}$ is a collection of clicked items where $L$ is the session length and $t_l$ is the id of the item the user clicked at position $l$. Given $s_i$, the session-based recommendation is to recommend the next most likely clicked item $v_{t_{L+1}}$. In practice, the recommender takes $s_i$ as input and predict the probability distributions of the the next item $p(v_{t_{L+1}}|s_i)$. Then the items with the top-K largest probability scores are recommended.


\subsection{Consecutive Intent Unit}
Current session-based recommendation methods consider each item separately and the higher level intent of a local session snippet are ignored. In this work, we not only learn the intent revealed by the individual items but also the combined intent from items in a consecutive fragment. We define $v_{j}^k$ as a consecutive intent unit in which the items are from a consecutive fragment started from $j$ and with length $k$, i.e., $v_{j}^k=(v_j,...v_{j+k-1})$. The length $k$ also represents the granularity level of the intent unit. An example is illustrated in Figure~\ref{fig:main} where a session $s=\{v_{1}, v_{2}, v_{1}, v_{3}, v_{2}, v_{1}, v_{3}, v_{4}\}$ is given. The items from the original sequence denoted by $v_1^1,...v_4^1$ compose the level-1 consecutive intent units. The session snippets $~(v_1,v_2),...,~(v_3,v_4)$ are level-2 consecutive intent units denoted by $v_1^2,...,v_5^2$ and  $~(v_1,v_2,v_1),...,~(v_1,v_3,v_4)$ are the level-3 consecutive intent units denoted by $v_1^3,...,v_5^3$.

Given session $s$, we assign a learnable embedding vector $e_j^1$ to represent the intent of the level-1 consecutive intent unit revealed by item $v_j$. For the high order level-$k$ consecutive intent units, we utilize a readout function $\mathcal{R}$ to fuse intents of each item in the corresponding intent unit to generate the representation. 
Specifically, the intent of a level-$k$ consecutive intent unit can be calculated by 

\begin{equation}
    e_{j}^{k}=\mathcal{R}(\{e_{j}^1,...,e_{j+k-1}^1\})
\end{equation}

We consider two kinds of readout functions i.e. set-based and sequence-based to generate representation for high level intent unit. The set-based readout functions e.g. MEAN, MAX etc. can extract order-invariant intent while the sequence-based readout functions e.g. Gate Recurrent Unit (GRU) can extract order-sensitive intent. In our method, we utilize both kinds of readout functions to extract the complete intent for high order intent unit. For $j$-th level-$k$ consecutive intent unit the final intent can be computed by


\begin{equation}
    e^{k}_{j}=e^{k,set}_{j}+e^{k,seq}_{j}
\label{eq:semantics}
\end{equation}

\noindent Where $e^{k,set}_{j}$ and $e^{k,seq}_{j}$ are corresponding to the order-invariant and order-sensitive intent respectively. 

\subsection{Multi-granularity Intent Heterogeneous Session Graph Construction}

In this section we introduce the construction of the \textit{Multi-granularity intent unit Heterogeneous Session Graph}~(MIHSG). The proposed MIHSG is composed by multiple subgraphs and each subgraph models the transitions between the intents of same level consecutive intent units. The graph constructed by the level-$k$ consecutive intent units is defined as the level-$k$ intent session graph. We first describe how to construct the level-$k$ intent session graph and then introduce the proposed MIHSG.


\subsubsection{Level-$k$ Intent Session Graph}

The level-$k$ session graph captures the spatial continuity of the user-item interactions. It is a directed graph $\mathcal{G}_s^{k}=(\mathcal{V}_s^{k},\mathcal{E}_s^{k})$ where each node represents the intent of a level-$k$ consecutive intent unit and each edge connects two consecutive adjacent level-$k$ consecutive intent units. Nodes have connection if they are adjacent in the session item sequence. The level-1 session graph captures the fine-grained intent transition between items. As the length of intent unit increases, the session graph contains higher level transition patterns between intent units.



\subsubsection{Multi-granularity intent unit Heterogeneous Session Graph}

We merge session graphs from different granularity levels to form a unified heterogeneous session graph i.e. MIHSG. Specifically, we introduce a special edge i.e. inter-granularity edge to link the high level session graphs and the level-$1$ consecutive session graph. It connects intent unit with its level-1 predecessor and successor.  The order of a MIHSG $k$ is the same as the largest granularity level. An example of MIHSG is given as illustrated in Figure~\ref{fig:framework}. In this graph, each node type represents an intent granularity level  and two kinds of edges are considered i.e. intra-granularity edge and inter-granularity edge. The intra-granularity edge denoted by $(v^{k},\text{intra-}k,v^{k})$ connects intents of the same granularity level as discribed in section 3.3.1 while the inter-granularity edge i.e. $(v^{1},\text{inter},v^{k})$ and $(v^{k},\text{inter},v^{1})$ models the intent transition between high level consecutive intent unit and their adjacent individual items. For example, in session $s=\{v_1,v_2,v_1,v_3\}$, two inter-granularity edges can be constructed. They are ~$(v_1,\text{inter},(v_2,v_1))$ and ~$((v_2,v_1),\text{inter},v_3)$. This enables capturing cross granularity level intent transition patterns. We do not consider inter-granularity edge between high granularity levels to avoid redundancy. 

\subsection{Session Representation Learning}

In this section, we introduce how to generate the representations for consecutive intent units and the whole session.

\subsubsection{Learning Representation of Consecutive intent units}

We employ the heterogeneous graph attention network~(HGAT) to learn the representation of each consecutive intent unit. 
Given an directed edge $(s,e,t)$ where $s$ and $t$ are the source and target intent units and $e$ is the edge.  Without loss of generality, $s$ and $t$ can be in any granularity levels. Specially, we denote $k_s$ and $k_t$ as the granularity levels of $s$ and $t$ respectively and denote $\phi_e$ as the edge type.  For a level-$k$ MIHSG, $k_s,k_t\in \{1,...,K\}$ and $\phi_e\in\{\text{inter},\text{intra-}1,...,\text{intra-}K\}$
. For each layer, we apply the bidirectional attention to aggregate the representations of the direct in-neighbors and out-neighbors. Given an in-neighbor set i.e. ${N}_{\phi_e}$, the aggregation mechanism to aggregate the in-neighbors is as following:
\vspace{-0.3cm}


\begin{small}
\begin{equation}
\begin{split}
    h_{t}^{(l+1)}&=\sum_{\phi_e}\sum_{s\in\mathcal{N}_{\phi_e}(t)} \alpha_{s}^{(l)} W^{(l)}_{\phi_e} h_s^{(l)} \\
    \alpha_{s}^{(l)}&=\mathop{\text{Softmax}}_{s}\left (\tilde{\alpha}_{s}^{(l)}\right ) \\
    \tilde{\alpha}_{s}^{(l)}&=\sigma\left (\vec{\textbf{a}}_{\phi_e}^{(l)^{T}} [W_{\phi_e}^{(l)}h_s^{(l)};W_{\phi_e}^{(l)} h_t^{(l)}] \right )
    \label{eq:hgat}
\end{split}
\end{equation}
\end{small}

\noindent Where $W^{(l)}_{\phi_e} \in \mathbb{R}^{d\times d}$ and ${\textbf{a}}^{(l)}_{\phi_e} \in {\mathbb{R}}^{2d\times 1}$ is the learnable projection weights which are not shared between different layers and edge types. $h_s^{(0)}=e_s$,$h_t^{(0)}=e_t$ is the initial representation of intent units $s$ and $t$. Following~\cite{velickovic2018graph}, we also adopt LeakyReLU activation function $\sigma(\cdot)$ to increase the non-linearity. 

We also adopt multi-way attention to stabilize the learning process. After getting the representation of each head, a readout function is applied to generate the output node representation.  

\begin{small}
\begin{equation}
\begin{split}
  h_{t}^{(l+1)}&=\mathop{\mathcal{R}}_{i=1,...,H}\left ( h_{t}^{(l+1),i}\right ) \\
  \label{eq:multihead}
  \end{split}
\end{equation}
\end{small}

\vspace{-0.4cm}

\noindent where $H$ is the number of attention heads and $i$ is head index. We empirically find element-wise max operation is consistently better than concatenation and mean operations.




As MIHSG is a directed graph, each node has context from both in-neighbors and out-neighbors. To extract context information from two directions, we apply the HGAT to aggregate the in-neighbors and out-neighbors information to generate node representations. Thus for each intent unit $v$, we get two embeddings  i.e. $\overrightarrow{h}^{(l+1)}_v$ and $\overleftarrow{h}^{(l+1)}_v$ for two directions respectively. Then the local representation of node $v$ is the summation of node embeddings from two directions, $\tilde{h}_v^{(l+1)}=\overrightarrow{h}^{(l+1)}_v + \overleftarrow{h}^{(l+1)}_v$. The final representation of node $v$ is the summation of the local representation and the mean of all embeddings in the session $h_{v}^{(l+1)}=\tilde{h}_v^{(l+1)}+\overline{h}_{u\in s}^{(l+1)}$.


\subsubsection{Learning Representation of Whole Session}

Current methods generate the user preferences by combining the representations of the last clicked individual item and the whole session. Since the high order granularity intents are ignored, the user preferences captured by these methods may be inaccurate or incomplete. To better utilize information of all granularity intents, we generate a separate session embedding for each level of consecutive intent units. 
As illustrated in Figure~\ref{fig:framework}, given a session $s_i$ and the corresponding consecutive intent unit embeddings $h^{k}_{i}\in\mathbb{R}^{d},i=1,...,n_k,k=1,...,K$, where $n_k$ is the number of intent units for level-$k$ and $K$ is the number of intent levels. For each level of consecutive intent units, we generate a local representation $z^{k}_l$ and a global representation $z^{k}_g$ to capture the user preferences. We take the last intent unit $h^{k}_{n_k}$ as $z^{k}_l$ and adopt a soft attention mechanism to generate $z^{k}_g$. To make session representation from each level captures complete user intent we compose embeddings of all intent units to generate a context set i.e. $C=\{h^{k}_{i}|i=i,...,n_k, k=1,...,K\}$ and denote $h_c 
\in C$ as one of the context embedding. Then we obtain $z^{k}_g$ by:
 
\begin{small}
 \begin{equation}
 \setlength{\abovedisplayskip}{2pt}
\setlength{\belowdisplayskip}{2pt}
   \begin{split}
     z^{k}_{g}=\sum_{c=1}^{|C|}\text{Softmax}_c(\gamma_{c}^{k})h_{c}
   \end{split}
   \label{eq:8}
 \end{equation}
 \end{small}

\noindent where the priority $\gamma_{c}^{k}$ is decided by the corresponding local representation $z^{k}_l$ and the contexts. Specifically, $\gamma_{c}^{k}$ is calculated by:

\begin{small}
\begin{equation}
  \gamma_{c}^{k}={W_0^{k}}^{T}\sigma(W_1^{k}h_c+W_2^{k}z_l^{k}+b^{k})
  \label{eq:9}
\end{equation}
\end{small}

\vspace{-0.3cm}

\noindent where $W_0^k\in \mathbb{R}^d$, $W_1^{k} \in \mathbb{R}^{d\times d}$, $W_2^{k} \in \mathbb{R}^{d\times d}$ are learnable parameters, $b^{k} \in \mathbb{R}^d$ is the bias and $\sigma(\cdot)$ is the sigmoid function. Since queries are different, the importance of each intent unit for different granularity levels are distinguished, which making a balance between capturing complete user intent and generating distinguishable level-$k$ session representation. We then combine the local representation and global representation to generate the user preferences for each level of consecutive intent units. The representations of user preferences for level-$k$ can be calculated by:

\begin{small}
\begin{equation}
  z_s^{k}=W_3^{k}[z_g^{k};z_l^{k}]
  \label{eq:10}
\end{equation}
\end{small}

\vspace{-0.3cm}

\noindent where $[\cdot]$ is the concatenation operation and $W_3^{k}$ is the projection matrix to transform $z^{k}$ from $\mathbb{R}^{2d}$ to $\mathbb{R}^d$.

\subsection{Intent Fusion Ranking and Optimization}

After generating session embeddings from different granularity levels i.e.  $z_s^{1},..., z_s^{k}, ...,z_s^{K}$, we propose to utilize Intent Fusion Ranking mechanism to capture comprehensive user preferences. Specifically, we first make a separate recommendation based on each level of intents and then fuse the results to make the final recommendation.

For each intent level $k$, we adopt the inner product to calculate the similarity between candidate item intent and the level-$k$ session embedding. Specifically, given a candidate item set $I=\{v_1, v_2,...v_{|I|}\}$, we can calculate the similarity between the session embedding $z_s^{k}$ and $e_{i}^{1}$ which is the intent of item $v_i$ as following:

\begin{small}
\begin{equation}
  y^{k}_i=\langle z_s^{k}, e_{i}^{1}\rangle
  \label{eq:11}
\end{equation}
\end{small}

\noindent where $\langle\cdot, \cdot\rangle$ is the inner product operate.


Inspired by~\cite{DBLP:conf/aaai/RenCLR0R19, DBLP:journals/corr/abs-2012-05422}, we also consider the repeat click behavior and exploration behavior. Specifically, we distinguish the inner session items from outer session items and apply the Softmax normalization separately as: 


\begin{small}
\begin{equation}
    \tilde{y}_{r_i}^{k} = \frac{\exp\left(y_{r_i}^{k}\right)}{\sum_{j=1}^{|R|}\exp\left(y_{r_j}^{k}\right)}, \ \ \ \ \ 
    \tilde{y}_{o_i}^{k} = \frac{\exp\left( y_{o_i}^{k}\right)}{\sum_{j=1}^{|O|}\exp\left( y_{o_j}^{k}\right )}
  \label{eq:12}
\end{equation}
\end{small}

\noindent where $R=\{r_1, ..., r_i, ..., r_{|R|}\}$ and $O=\{o_1, ..., o_i, ..., o_{|O|}\}$ represent the inner session items and outer session items respectively and we have $|R| + |O| = |I|$. $\tilde{y}_{r_i}^{k}\in\mathbb{R}$ and $\tilde{y}_{o_i}^{k}\in\mathbb{R}$ are the corresponding normalized probabilities. By separate normalization, we distinguish the repeat click behavior and exploration behavior.


Then, we use a discriminator to re-weight the item scores to balance the focus between repeat click and exploring click. Then the scores $\textbf{y}$ is the combination of the two parts, by:

\begin{small}
\begin{equation}
  \begin{split}
   \textbf{y}^{k}&=[\beta_{r}^{k}\tilde{\textbf{y}}_{r}^{k};\beta_{o}^{k}\tilde{\textbf{y}}_{o}^{k}] \\
    \beta_{r}^{k},\beta_{o}^{k}&=\text{Softmax}\left (W_1^T\sigma\left (W_2 z_s^{k}\right ) \right)
  \end{split}
  \label{eq:13}
\end{equation}
\end{small}

\noindent where $[;]$ is the concatenation operation, $\tilde{\textbf{y}}_r^k\in\mathbb{R}^{|R|}$ and $\tilde{\textbf{y}}_o^k\in\mathbb{R}^{|O|}$ are probability distributions for inner session items and outer sessions items respectively and we have $\tilde{y}_{r_i}^{k}\in\tilde{\textbf{y}}_r^k$, $\tilde{y}_{o_i}^{k}\in\tilde{\textbf{y}}_o^k$. $W_1\in\mathbb{R}^{d\times 2}$ and $W_2\in\mathbb{R}^{d\times d}$ are the learnable projection matrix, $\sigma$ is the sigmoid function. We denote this strategy as Repeat-Explore Normalization~(RENorm). 


We propose the \textit{Intent Fusion Ranking}~(IFR) to fuse the recommendation results predicted by all granularity levels of intents. Specially, we introduce a weighted summation operator to fuse the probability distributions generated by all levels of intents to generate the final probability distribution $\hat{\textbf{y}}$.


\begin{small}
\begin{equation}
    \hat{\textbf{y}}=\sum_{k=1}^K\hat{\alpha}^{k}\textbf{y}^{k} , \ \ \ \ 
   \hat{\alpha}^{k}=\frac{\exp{(\alpha^{k})}}{\sum_{k=1}^K\exp{(\alpha^{k})}}
  \label{eq:14}
\end{equation}
\end{small}

\noindent where $\alpha^{k}$ is the learnable factor for each probability distribution i.e. $\textbf{y}^{k}$ and $\hat{\alpha}^{k}$ is the normalized weight.

We adopt cross-entropy as the optimization objective to learn the parameters and the loss function is:

\begin{equation}
  L(\hat{\textbf{y}})=-\sum_{i=1}^{|I|}y_i\log(\hat{y}_i)+(1-y_i)\log(1-\hat{y}_i)
  \label{eq:15}
\end{equation}

\noindent where $y_i\in \textbf{y}$ reflects the appearance of an item in the one-hot
encoding vector of the ground truth, i.e., $y_i=1$ if the $i$-th item is
the target item of the given session; otherwise, $y_i=0$. We also add $l_2$ norm on the item embeddings to prevent popularity-bias phenomenon~\cite{DBLP:journals/corr/abs-1909-04276} and use scaled softmax with scaled factor 12 at the normalization stage Eq.~(\ref{eq:12}) to prevent over smoothing. In addition, following~\cite{DBLP:conf/cikm/PanCCCR20}, we apply the Back-Propagation Through Time (BPTT) algorithm~\cite{werbos1990backpropagation} to train the MSGIFSR model. 

    



\section{Experiments}
In this section, we first describe the experimental settings including the datasets, baslines, and evaluation metrics. Then we do detailed analyses for the experimental results.. 


\subsection{Datasets}

 We conduct the experiments on four real-world datasets which are commonly used in the literature of session-based recommendation.
\begin{itemize}
    \item \textit{Diginetica} is a personalized e-commerce search challenge dataset provided in CIKM Cup 2016. The dataset contains transition history which is suitable for session-based recommendation. Following~\cite{DBLP:conf/aaai/WuT0WXT19,DBLP:conf/aaai/RenCLR0R19,DBLP:conf/sigir/Yu0LWWT20,DBLP:conf/kdd/ChenW20,DBLP:conf/kdd/LiuZMZ18,DBLP:conf/cikm/LiRCRLM17}, we use the sessions in the last week for test.
    \item \textit{Gowalla} is a check-in dataset that is widely used for point-of-interest recommendation. Following~\cite{,DBLP:conf/kdd/ChenW20,DBLP:conf/kdd/GuoYWCZH19,DBLP:conf/wsdm/TangW18}, we kept the top 30,000 most popular locations, and grouped users’ check-in records into disjoint sessions by splitting intervals between adjacent records that are longer than 1 day. The last 20\% of the sessions were used as the test set.
    \item \textit{Last.Fm} is widely used in many recommendation tasks. The music artist recommendation is used as the task. Following~\cite{DBLP:conf/kdd/ChenW20,DBLP:conf/aaai/RenCLR0R19,DBLP:conf/kdd/GuoYWCZH19}, we kept the top 40,000 most popular artists and set the splitting interval to 8 hours. Similar to Gowalla, the most recent 20\% of the sessions were used as the test set.
     \item \textit{Yoochoose} is a dataset that is obtained from the RecSys Challenge 2015, which contains a stream of user clicks on an e-commerce website within 6 months. We use the typical  method in~\cite{DBLP:conf/aaai/WuT0WXT19,DBLP:conf/aaai/RenCLR0R19,DBLP:conf/sigir/Yu0LWWT20,DBLP:conf/kdd/ChenW20,DBLP:conf/kdd/LiuZMZ18,DBLP:conf/cikm/LiRCRLM17} to split the dataset and use the 1/64 and 1/4 subsample of all training sessions as the training set.
    
\end{itemize}

\noindent Following~\cite{DBLP:conf/cikm/QiuLHY19,DBLP:conf/aaai/WuT0WXT19,DBLP:conf/sigir/PanCLR20a,DBLP:conf/cikm/LiRCRLM17,DBLP:conf/aaai/RenCLR0R19,DBLP:conf/kdd/LiuZMZ18,DBLP:conf/sigir/Yu0LWWT20,DBLP:conf/kdd/ChenW20}, we filter out sessions of length 1 and items appearing less than 5 times. We adopt the data augmentation method described in in~\cite{DBLP:conf/cikm/LiRCRLM17,DBLP:conf/aaai/RenCLR0R19,DBLP:conf/kdd/LiuZMZ18} to process the dataset. The statistics of these datasets are described in Table \ref{tab:dataset}.
\vspace{-0.3cm}
\begin{table}[ht]
    \centering
    \caption{Statistics of datasets}
    \resizebox{\linewidth}{!}{
    \begin{tabular}{cccccc}
        \toprule
                & Diginetica & Gowalla & Last.FM & yoochoose1/64 & yoochoose1/4\\
        \midrule
        \#clicks &981,620 &1,122,788 & 3,835,706 & 557,248 & 8,326,407 \\
        \#train sessions  & 716,835 & 675,561 & 2,837,644 & 369,859 & 5,917,745 \\
        \#test sessions & 60,194 & 155,332 & 672,519 & 55,898 & 55,898 \\
        \#items & 42,596 & 29,510 &  38,615 & 16,766 & 29,618 \\
        $\#\text{length} \le 5$ & 537,546 & 627,100 & 1,136,909  & 289,490 & 4,234,915\\
        $\#\text{length} > 5$ & 239,483 & 203,793 &  2,373,254 & 136,267 & 1,738,734\\
        Average length & 4.80 & 4.32 & 9.16 & 6.16 &  5.71 \\
        \bottomrule
    \end{tabular}}
    \label{tab:dataset}
\end{table}
\vspace{-0.5cm}

\subsection{Baselines}

We consider three kinds of methods as our baselines: the conventional Nearest Neighbors (NN) based methods, Neural-based sequence methods and GNN based methods. The description of the baselines are as follows:

\begin{itemize}
    \item \textbf{Item-KNN}~\cite{10.1145/1864708.1864770}  is a NN based method which recommends items that are similar to the previous items in current session and cosine similarity is adopted to measure the similarity between two items.
    \item \textbf{GRU4Rec}~\cite{DBLP:journals/corr/HidasiKBT15} employs gated recurrent unit~\cite{cho-etal-2014-learning} to model the sequential behaviour of items in a session.
    \item \textbf{NARM}~\cite{DBLP:conf/cikm/LiRCRLM17} adopts GRU to extract sequence information and employs attention mechanism to capture user preferences.
    \item \textbf{SR-GNN}~\cite{DBLP:conf/aaai/WuT0WXT19} uses gated graph neural networks to capture the transition relations between items.
    \item \textbf{GC-SAN}~\cite{DBLP:conf/ijcai/XuZLSXZFZ19} applies a self-attention layer after the graph neural network module to integrate contextual information.
    \item \textbf{NISER+}~\cite{DBLP:journals/corr/abs-1909-04276} utilizes dropout and $l_2$ norm to alleviate over-fitting and long-tail effect.
    \item \textbf{SGNN-HN}~\cite{DBLP:conf/cikm/PanCCCR20} uses highway network to reduce over-smooth and long-range dependency problems.
    \item \textbf{LESSR}~\cite{DBLP:conf/kdd/ChenW20} introduces two kind of session graphs to solve the information loss and long-range dependency problem.
    \item \textbf{SHARE}~\cite{wang2021session} proposes utilizing hypergraph and attention network to exploit the contextual windows to model session-wise item representations.
\end{itemize}

\vspace{-0.3cm}
\subsection{Experimental Setup}

For all methods, we employ grid search to find the best hyper-parameters. We random split 10\% samples from the training set as the validation set and the hyper-parameters which achieve best performance on the validation set are selected. Specifically, for GNN-based models, the number of layers is searched in $\{1,2,3,4,5\}$. For the proposed MSGIFSR, we also search the best intent unit granularity level $K$ in $\{1,2,3,4,5,6,7\}$. We use Adam to optimize the model while set learning rate and weight decay to $1e^{-3}$ and $5e^{-4}$ respectively. We follow~\cite{DBLP:conf/aaai/WuT0WXT19,DBLP:conf/cikm/PanCCCR20,DBLP:journals/corr/abs-1909-04276} to decay the learning rate every 3 epochs with a 0.1 rate. We fix the embedding dimension to 256 and the batch size to 512. For all models, we compare the HR@20 (Hit Rate) and MRR@20 (Average Reciprocal Ranking) metrics. 

\begin{table*}[htpb]
\setlength{\abovedisplayskip}{1pt}
\setlength{\belowdisplayskip}{1pt}
\centering
  \caption{Results(\%) of main experiments. $\ast$ denotes a significant improvement of MSGIFSR over the best baseline using a paired $t$-test ($p$ < 0.01).}
  \label{fig:main}
  \resizebox{.8\linewidth}{!}{
  \begin{tabular}{ccccccccccc}
    \toprule
    \multirow{2}*{Model} & \multicolumn{2}{c}{Diginetica} & \multicolumn{2}{c}{Gowalla} & \multicolumn{2}{c}{Last.FM} & \multicolumn{2}{c}{Yoochoose 1/64} & \multicolumn{2}{c}{Yoochoose 1/4} \\
               ~ & HR@20 & MRR@20 & HR@20 & MRR@20 & HR@20 & MRR@20 & HR@20 & MRR@20 & HR@20 & MRR@20 \\
    \midrule
    Item-KNN & 39.51 & 11.22 & 38.60 & 16.66 & 14.90 & 4.04 & 51.60 &  21.81 &  52.31 & 21.70 \\
    GRU4Rec & 42.55 & 12.67 & 39.55 & 16.99 & 22.13 & 7.15 & 61.38 & 23.86 & 71.40 & 29.95 \\
    NARM   & 52.89 & 16.84 & 52.24 & 25.13 & 23.09 & 7.90 & 70.40 & 30.17 & 72.11 & 30.62  \\
    SR-GNN & 53.44 & 17.31 & 53.24 & 26.03 & 23.85 & 8.23 &  70.85 & 30.71  & 72.55 & 32.09 \\
    GC-SAN & 54.77 & 18.57 & 53.66 & 25.69 & 22.64 & 8.42 &  71.44 & 31.65 & 71.77 & 32.10 \\
    NISER+ & $\textbf{56.56}$ & 19.38 & $\textbf{55.33}$ & 26.67 & 24.76 & 9.02 & 72.08 & 32.60 & 73.52 & 32.63 \\
    SGNN-HN & 55.67 & \textbf{19.45} & 55.28 & \textbf{27.58} & \textbf{25.07} & $\textbf{9.40}$ & \textbf{72.13} & \textbf{32.60} & \textbf{73.52} & 32.63 \\
    LESSR  & 51.71 & 18.15 & 51.34 & 25.49 & 23.37 & 9.01 & 70.59 & 31.46 &  72.67 & \textbf{33.12}  \\
    SHARE  & 55.87 & 18.83 & 54.97 & 26.16 & 23.83 & 7.49 & 71.97 & 31.85 & 73.26 & 31.89 \\
    \midrule
    MSGIFSR& $\textbf{57.11}^{\ast}$ & $\textbf{20.05}^{\ast}$ & $\textbf{56.64}^{\ast}$ & $\textbf{29.02}^{\ast}$ & $\textbf{27.63}^{\ast}$ & $\textbf{10.86}^{\ast}$ & $\textbf{73.13}^{\ast}$ & $\textbf{33.50}^{\ast}$ & $\textbf{74.01}^{\ast}$ & $\textbf{33.74}^{\ast}$ \\
    \midrule 
    \% Gain & $\textbf{0.97\%}$  & $\textbf{3.08\%}$  & $\textbf{2.37\%}$ & $\textbf{5.22\%}$ & $\textbf{10.21\%}$ & $\textbf{15.53\%}$ & $\textbf{1.39\%}$ & $\textbf{2.76\%}$ & $\textbf{0.67\%}$ & $\textbf{1.87\%}$  \\
    \bottomrule
  \end{tabular}}
\end{table*}

\vspace{-0.2cm}

\subsection{Performance Comparison}

In this section, we compare our methods with the state-of-the-art baselines to validate the effectiveness. The empirical results of all methods are shown in Table~\ref{fig:main}. From the results, we have the following important observations: 



    


Firstly, neural network based methods are significantly better than conventional methods~(e.g. Item-KNN), proving the neural network based models are capable of capturing complex sequential patterns for making recommendations. However, among them, the GRU4Rec achieves inferior performance and the reason is that in GRU4Rec only sequential information is leveraged. In contrast, NARM achieves competitive performance because it exploits both of the sequential information and global preferences. Moreover, we find that GNN-based models~\cite{DBLP:conf/cikm/PanCCCR20, DBLP:conf/aaai/WuT0WXT19, DBLP:journals/corr/abs-1909-04276, DBLP:conf/ijcai/XuZLSXZFZ19, DBLP:conf/kdd/ChenW20,wang2021session} are better than previous methods. The GNN-based methods demonstrate the superiority of session graphs in representing transition relationships between different items. In current GNN based methods, SGNN-HN achieves better performance since it introduces a star node to help filter out irrelevant items and adopt high-way network to relieve the over-fitting problem.

Secondly, our method MSGIFSR outperforms all baselines by a large margin, indicating that the session-based recommendation can benefit from our proposed consecutive intent units. We attribute such significant gain to the following reasons: 1) the proposed MIHSG can exploit the intents from multiple granularity levels and model the complex transitions between different user intents; 2) the proposed IFR module can integrate the intents of different granularity levels to capture more accurate and complete user preferences. In addition, we find that MSGIFSR has larger gains on datasets which contain relative long sessions e.g. \textit{Last.FM}. The reason is that MSGIFSR captures the user preferences in a hierarchy way i.e. low level of intent unit to capture the fine grained intents while the high level intent unit to capture complex intents. Such a design can help propagate information between different intents more effectively and efficiently. We also notice that our proposed methods obtain relatively larger gains on \textit{Gowalla} and \textit{Yoochoose1/64}. From the statistics, we can learn that \textit{Gowalla} and \textit{Yoochoose1/64} have relative small number of items and sessions. In such situation, our proposed high level consecutive intent units can help extract richer user intent, thus boosting the recommendation performance.

\vspace{-0.3cm}
\begin{table}[htpb]
\Huge
\centering
  \caption{Results(\%) of ablation experiments.}
  \label{tab:ablation}
  \resizebox{\linewidth}{!}{
  \begin{tabular}{lcccccccccc}
    \toprule
    \multirow{2}*{Model} & \multicolumn{2}{c}{Diginetica} & \multicolumn{2}{c}{Gowalla} & \multicolumn{2}{c}{Last.Fm} & \multicolumn{2}{c}{Yoochoose1/64}  \\
               ~ & HR@20 & MRR@20 & HR@20 & MRR@20 & HR@20 & MRR@20 & HR@20 & MRR@20  \\
    \midrule
    MSGIFSR  & \textbf{57.11} & \textbf{20.05} & \textbf{56.64} & \textbf{29.02} & \textbf{27.63} & \textbf{10.86} & \textbf{73.13} & \textbf{33.50}  \\
    \midrule
\textemdash intra-E & 56.91 & 19.72 & 56.54 & 28.84 & 27.55 & 9.32 & 72.94 & 33.38  \\
\textemdash inter-E & 56.93 & 19.96 & 56.61 & 28.93 & 27.60 & 10.78 & 73.00 & 33.43 \\
\textemdash MIHSG & 56.42 & 19.64 & 55.38 & 28.12 & 26.47 & 9.23 & 72.47 & 32.68  \\
    \textemdash IFR & 56.12 & 19.37 & 55.45 & 28.29 & 26.96 & 10.61 & 72.02 & 32.58  \\
    \textemdash RENorm & 56.86 & 19.48 & 56.35 & 27.73 & 27.39 & 9.59 & 72.71 & 32.27  \\

    \bottomrule
  \end{tabular}}
\end{table}
\vspace{-0.5cm}


\subsection{Ablation Study}

In this section, we conduct ablation study to evaluate the contribution of each key components. Specifically, in each experiment we remove one of the components and the performance change indicates the importance of the removed component. The results are shown in Table~\ref{tab:ablation}.

\subsubsection{Effectiveness of Multi-granularity Intent Heterogeneous Session Graph}

To study the impact of MIHSG, we train models without intra-granularity edges~(intra-E), without inter-granularity edges~(inter-E), and without MIHSG~(without both types of edges) respectively and compare the performance. From Table~\ref{tab:ablation}, we find that either removing intra-granularity edges or inter-granularity edges harms HR@20 and MRR@20. It indicates that both kinds of transition relationships are important to capture the sequential information. By removing MIHSG, the transition information is completely lost and the context awareness is largely restricted since the sequence info has been removed. Meanwhile, by doing so, the propagation path for long dependency is cut off, which further aggravates the information loss. Besides, removing intra-granularity edges has a larger impact than removing inter-granularity edges. This is because the transitions between the intent units within the same granularity serve as the foundation of the user preference description, and the inter-granularity edges can also benefit from it.

\subsubsection{Impact of Intent Fusion Ranking Module and RENorm}

The intent fusion ranking module composes recommendation results from all intent granularity levels. To study the effectiveness of IFR, we train a reduced model only taking the results from the level-$1$ intent granularity to make recommendations. As shown in Table \ref{tab:ablation}, both of HR@20 and MRR@20 decays significantly, which means that information from higher intent granularity levels is complementary to the original level-$1$ intent granularity item sequence. We also train another reduced model without RENorm and find that worse MRR@20 are reported on all four datasets. It indicates that distinguishing repeat click behaviors and exploration behaviors is beneficial. By comparing the effectiveness of IFR module and RENorm module, it is observed that the IFR module brings larger gains on HR@20 while RENorm module impacts the MRR@20 more. An explanation is that the IFR module can help to recommend more correct items which favor HR and the RENorm aims to obtain strictly correct order, which leads to better MRR.

\subsection{In-depth Analysis}

\subsubsection{Performance on Different Session Lengths}

\begin{figure}[h]
\setlength{\abovecaptionskip}{0.1cm}
  \begin{center}
  \subfigure[Results on HR@20]{\includegraphics[width=.7\linewidth]{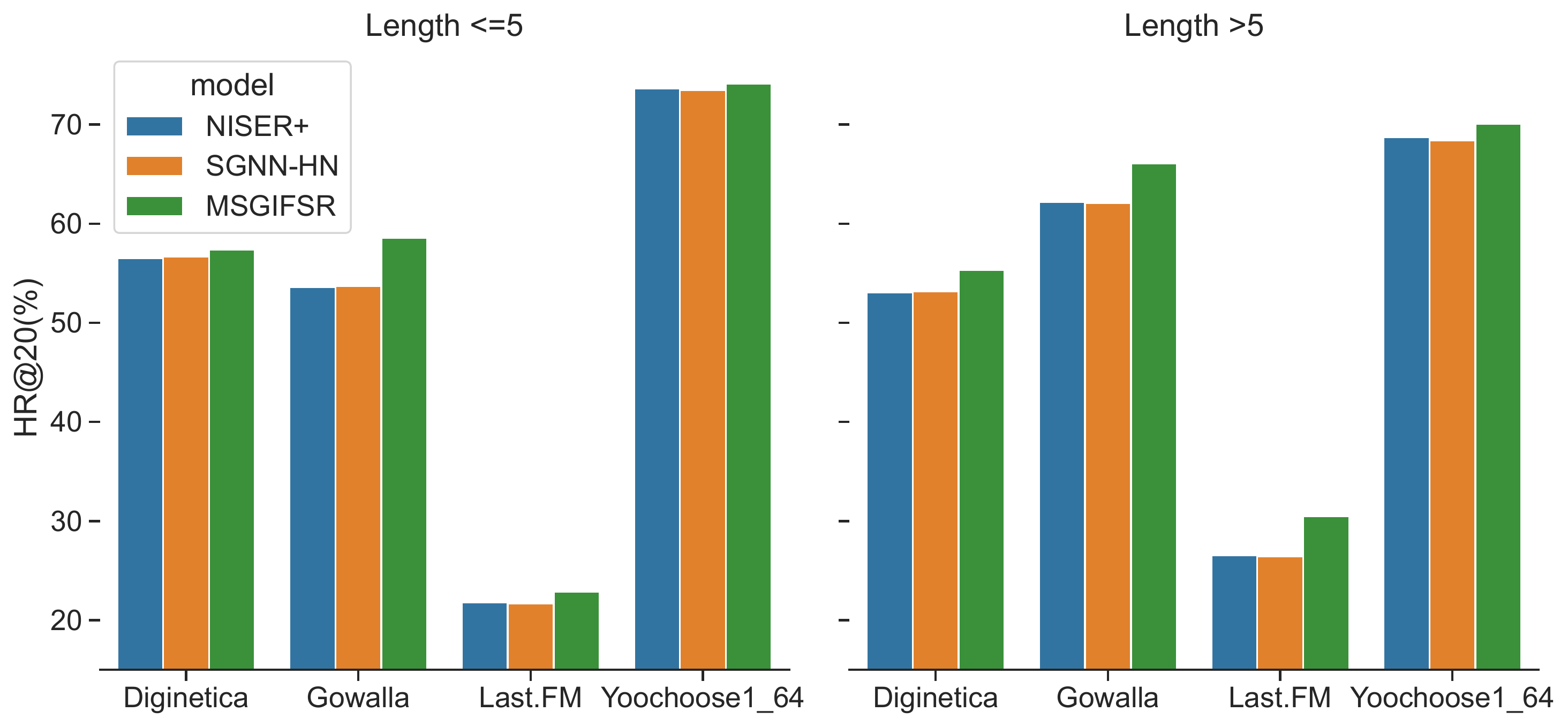}}
  \subfigure[Results on MRR@20]{\includegraphics[width=.7\linewidth]{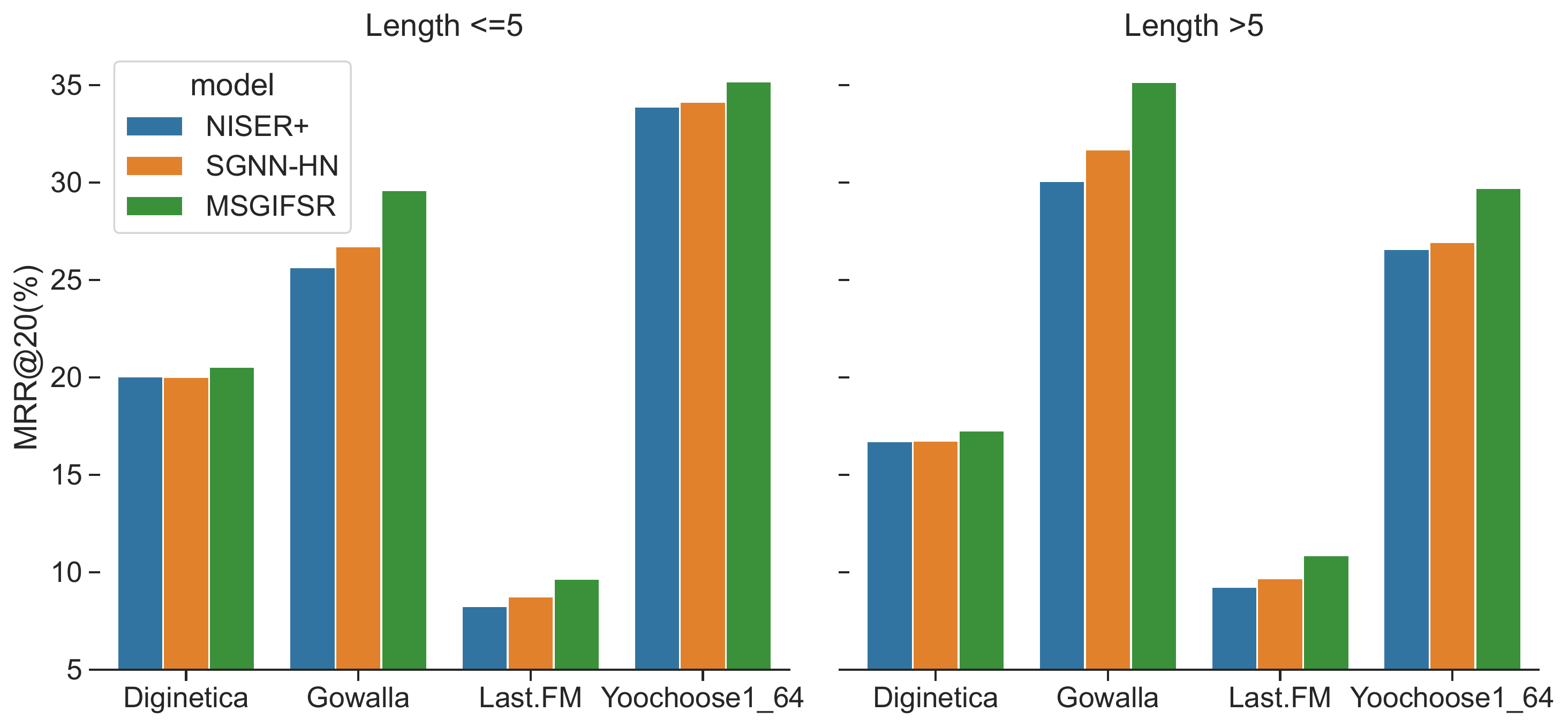}}
    \end{center}
  \caption{Comparison on different lengths of sessions.} 
  \label{fig:length}
\end{figure}

We study relation of the performance and session lengths. We firstly split the test set into long sessions and short sessions. Following~\cite{DBLP:conf/aaai/WuT0WXT19, DBLP:conf/cikm/PanCCCR20}, sessions with length larger than 5 are long sessions while remain sessions are short ones. After that, we compare our method, i.e., MSGIFSR with intent unit granularity level 3 and \textit{Mean-GRU} readout function with two state-of-the-art baselines NISER+ and SGNN-HN, and report the result on the four datasets. All GNN layers are set to 1 for fair comparisons. The results are shown in Figure~\ref{fig:length}. 

From the figure, we find that NISER+ performs worse than SRGNN-HN. It is because SRGNN-HN applies star node to filter out irrelevant items. Among the three datasets, MSGIFSR achieves the best result on both long and short sessions, indicating the effectiveness of MSGIFSR.
Moreover, we observe that the baseline methods suffer from the long-range dependency problem, especially for the long sessions, and the relative gains of MSGIFSR to baselines in long sessions are larger than in short ones. 
The reason is that the proposed high-level intent unit granularity can model more accurate user intent from a combined perspective, i.e., the overall meaning of a group of consecutive items. And, the information of the former intent units can be efficiently transported to the later ones through MIHSG. Thus, the long-dependency problem can be well handled.

\vspace{-0.5cm}

\subsubsection{Results on Different intent unit Granularity}

We test the result with respect to intent granularity level on four datasets. The results are shown in Figure~\ref{fig:levels}. We use the 1-layer HGAT network to avoid the influence of high-order information propagation. We find that MSGIFSR with higher level intent granularity is consistently better than that with only single items. For \textit{Diginetica}, in the beginning, the performance of MSGIFSR improves as higher level of intent granularity is incorporated into the MIHSG. Then the performance gain becomes stable as the level of granularity increases. We observe similar phenomenon in \textit{Yoochoose}, \textit{Gowalla}. For \textit{Last.FM}, the performance of MRR@20 has a drop when intent granularity level increase to 4 and 5 and then becomes better for $L=6$. It indicates that for the datasets with a long average session length, we need to incorporate higher-level granularity of intent to capture the user preference. We also find that as the granularity goes coarser, the performance becomes stable. That is because sessions are usually not so long and keep increasing the granularity level brings no more useful information.

\subsubsection{Comparisons with Hypergraph}

\begin{figure}[htbp]
\setlength{\abovecaptionskip}{0.1cm}
  \begin{center}
  \subfigure[Diginetica]{\includegraphics[width=.45\linewidth]{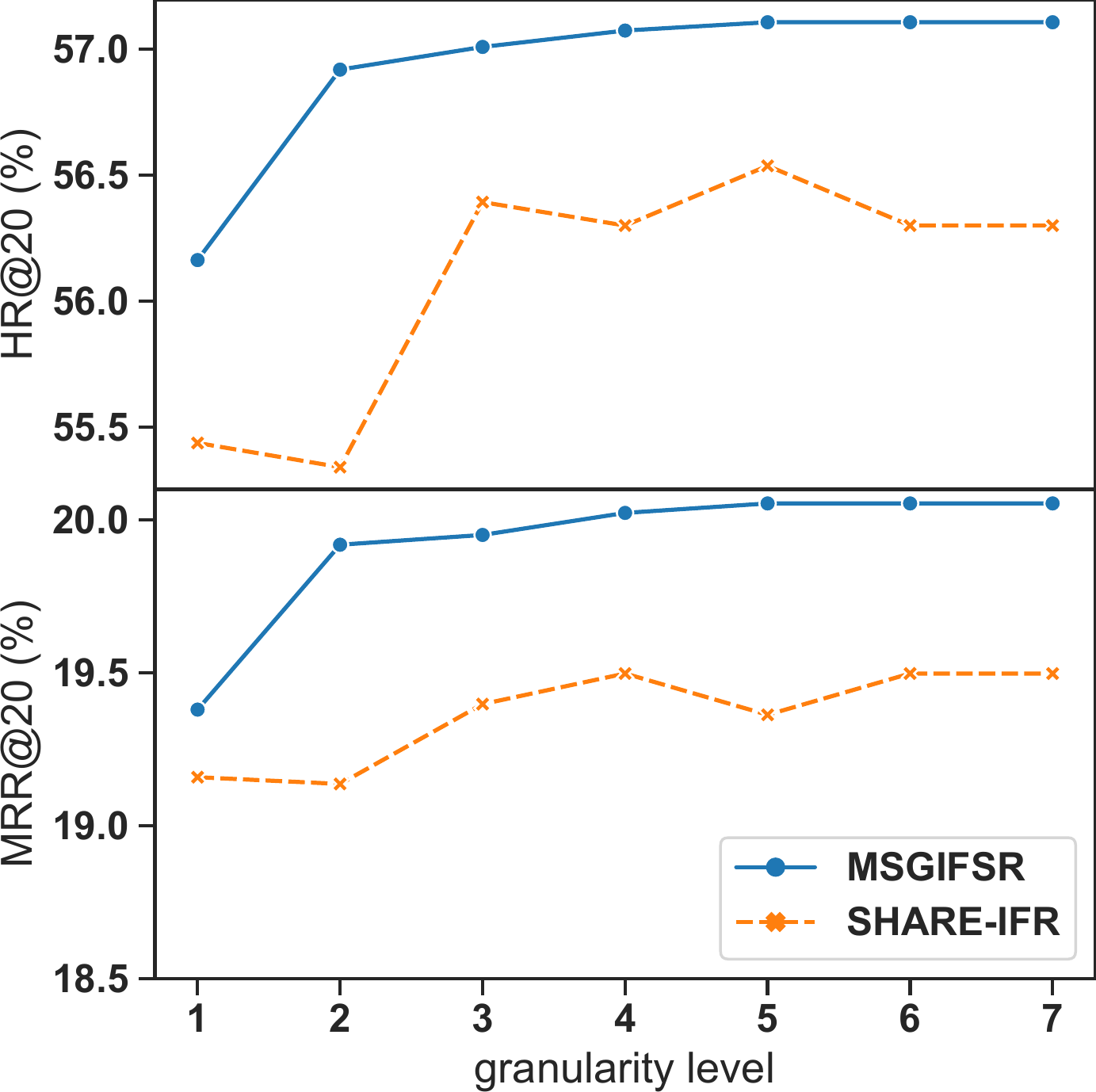}}
  \subfigure[Gowalla]{\includegraphics[width=.45\linewidth]{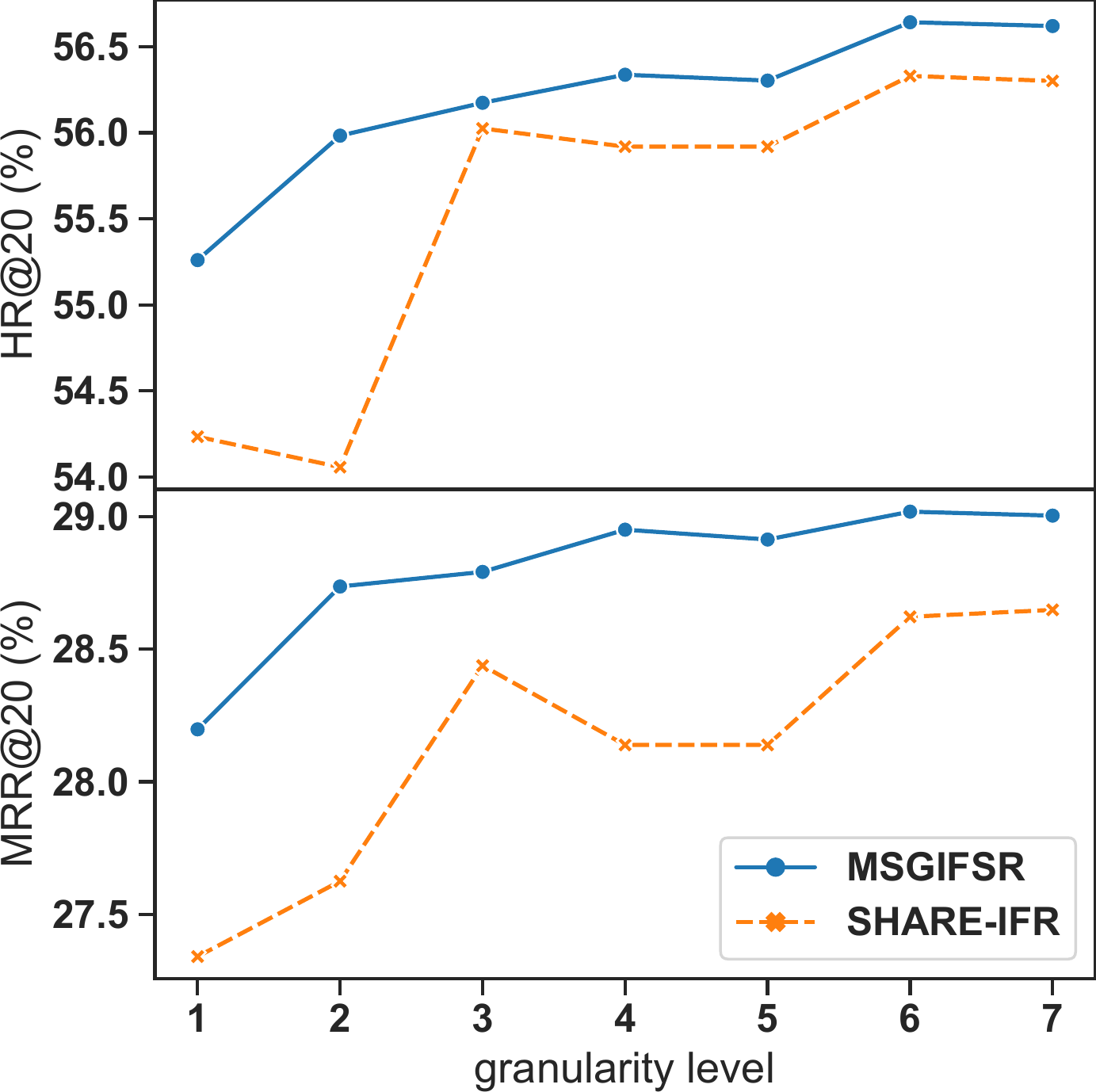}}
   \subfigure[Last.FM]{\includegraphics[width=.45\linewidth]{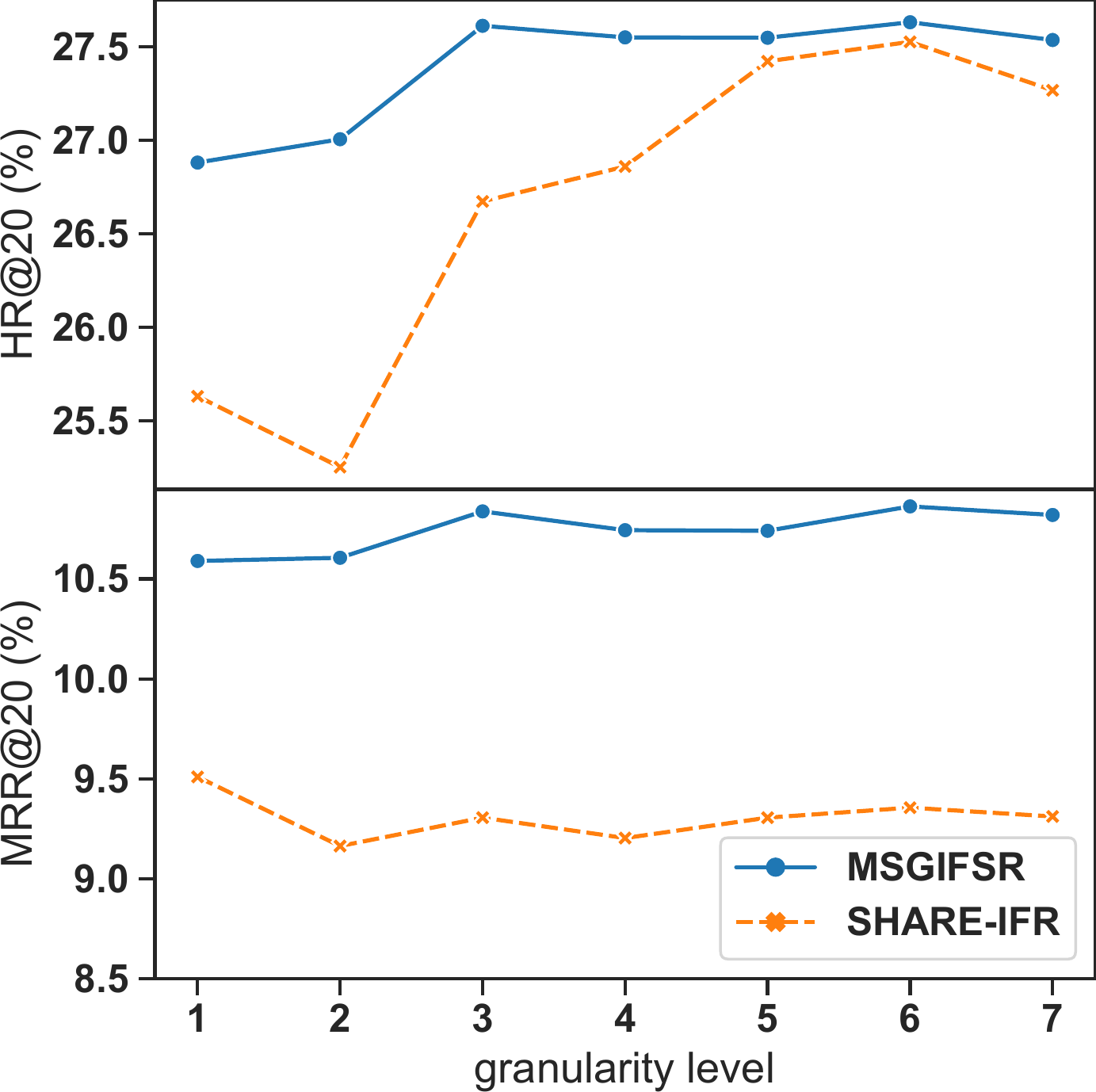}}
   \subfigure[Yoochoose1/64]{\includegraphics[width=.45\linewidth]{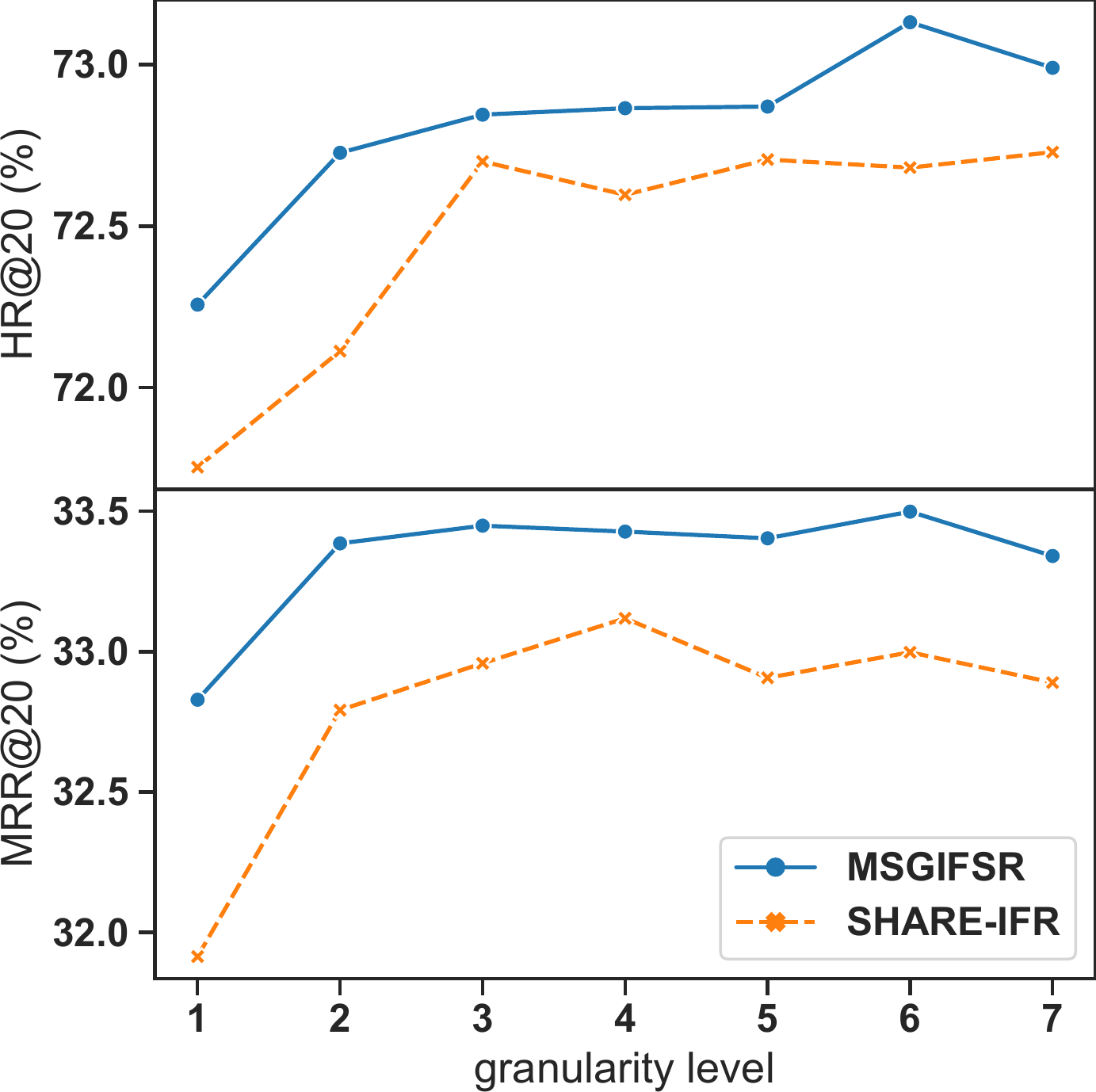}}
  
\end{center}
\caption{Impact of intent unit Granularity Levels}
  \label{fig:levels}
\end{figure}

We replace MIHSG with other topological structures like hypergraph~\cite{wang2021session} to test the effectiveness of MIHSG. We then apply a hypergraph attention neural network to perform message passing operations on such hypergraph while keeping the remaining settings the same as our model. We denote the model SHARE-IFR since it utilizes the proposed IFR module for fairness. The representations of hyperedges are considered as high-level granularity intent. The intent-granularity levels are among 1 to 7. The results are shown in Figure~\ref{fig:levels}. We find that MIHSG performs better than hypergraph across different granularity levels. It is because hypergraph only considers the group intent information of items and ignores the sequential behavior of them, which may cause information loss problem. In contrast, MIHSG can capture such information by using a recurrent neural network like \textit{GRU}. Moreover, we also find that SHARE-IFR is not stable as the intent granularity level increases. An explanation is that the hypergraph attention neural network conducts message passing by two steps, i.e., from nodes to hyperedges and then from hyperedges to nodes. This makes information transported between two hyperedges inefficiently and destabilizes the performance when multiple hyperedges are involved. While heterogeneous GAT makes message passing for nodes with different intent granularity levels directly, hence avoid such phenomenon.
\vspace{-0.25cm}
\vspace{-0.6cm}



\section{Conclusion}
In this paper, we study the session-based recommendation problem and proposed MSGIFSR, a novel model that extracting different in-session intent granularity information. It is observed that varied consecutive combined intent granularity provides richer user preference, and our proposed MIHSG successfully captures complex preference transition relationships among multi-level consecutive intent units and long-range dependencies by modeling intra- and inter-granularity intent unit edges. In addition, the intent unit encoder mechanism considering both the order-variant and order-invariant relations of a group of items which are good at representing intent unit meanings. Last but not least, the ablation study demonstrates the \textit{Intent Fusion Ranking} module successfully incorporates recommendation results from all intent unit levels. We will extend our work in a streaming session-based setting and reduce the time complexity and popularity bias for recommendations in the future work. 

\noindent \textbf{Acknowledgments.} Jiayan Guo and Yan Zhang are supported by National Key Research and Development Program of China under Grant No. 2018AAA0101902, and NSFC under Grant No. 61532001.

\newpage

\bibliographystyle{ACM-Reference-Format}
\balance
\bibliography{main.bbl}

\newpage
\appendix
\section{Appendices}

\subsection{Hyper-parameter Study}


We study the effects of different readout functions when extracting high-level granularity of intent from a group of items. Specifically, we compare order-invariant readout functions~(\textit{MEAN}, \textit{MAX}), order-variant readout functions~(\textit{GRU}) and the combination of both operations~(sum the outputs). The intent granularity length is set to 3 in all experiments. As shown in Figure~\ref{tab:readout}, for \textit{Diginetica} and \textit{Gowalla} the readout function with both order-variant operation and order-invariant operation achieves the best results. However, for \textit{Yoochoose1/64}, \textit{GRU} achieves the best result, indicating that sequential behavior is critical for predicting the next item. As for \textit{Last.FM}, \textit{MAX}+\textit{GRU} achieves the best in HR@20, since max pooling can filter out irrelevant items and \textit{GRU} preserves sequential information thus help make recommendation on long sessions. 

\begin{table}[ht]
\Huge
    \centering
    \caption{Impact of Readout function}
    \label{tab:readout}
    \resizebox{\linewidth}{!}{
    \begin{tabular}{ccccccccc}
        \toprule
                & \multicolumn{2}{c}{Diginetica} & \multicolumn{2}{c}{Gowalla} & \multicolumn{2}{c}{Last.FM} & \multicolumn{2}{c}{Yoochoose1/64} \\
                ~ & HR@20 & MRR@20 & HR@20 & MRR@20 & HR@20 & MRR@20 & HR@20 & MRR@20 \\
        \midrule
        MEAN & 56.87 & 19.91 & 55.89 & 28.66 & 26.95 & 10.46 & 72.73 & 33.35 \\
        
        MAX  & 56.82 & 19.84 & 55.88 & 28.57 & 26.99 & 10.55 & 72.60 & 33.38  \\
        
        
        GRU & 56.81 & 19.94 & 56.03 & 28.75 & 26.95 & 10.60 & \textbf{72.85} & \textbf{33.48} \\
        
        MEAN+GRU  & 56.83 & 19.94 & \textbf{56.07} & \textbf{28.76} & 26.99 & 10.63 & 72.80 & 33.33 \\ 
        MAX+GRU  & \textbf{56.90} & \textbf{20.03} & 56.00 & 28.74 & \textbf{27.01} & \textbf{10.65} & 72.80 & 33.42  \\
        \bottomrule
    \end{tabular}}
\end{table}

\begin{figure}[hb!]
  \begin{center}
  \subfigure[Diginetica]{\includegraphics[width=.4\linewidth]{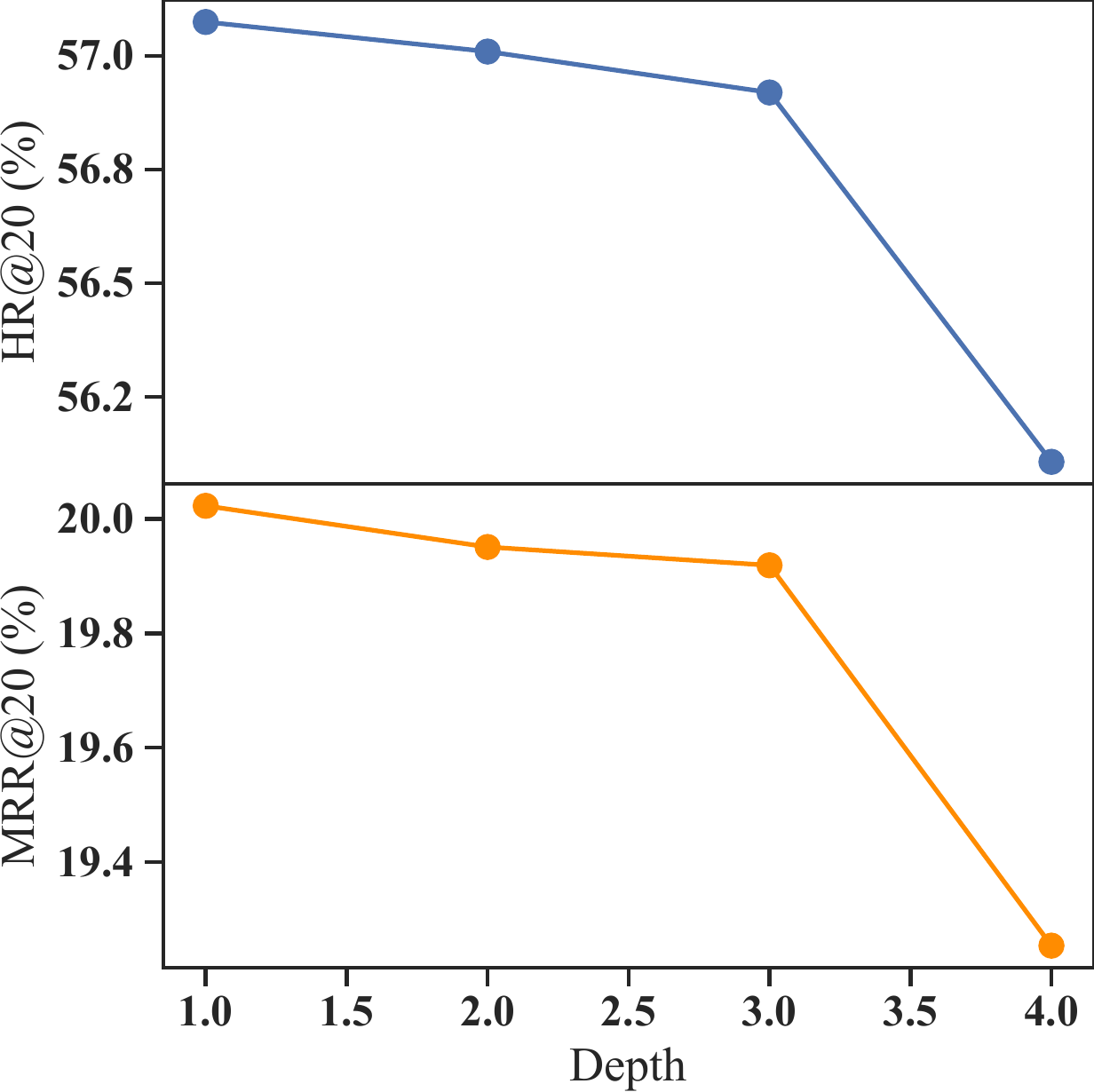}}
  \subfigure[Gowalla]{\includegraphics[width=.4\linewidth]{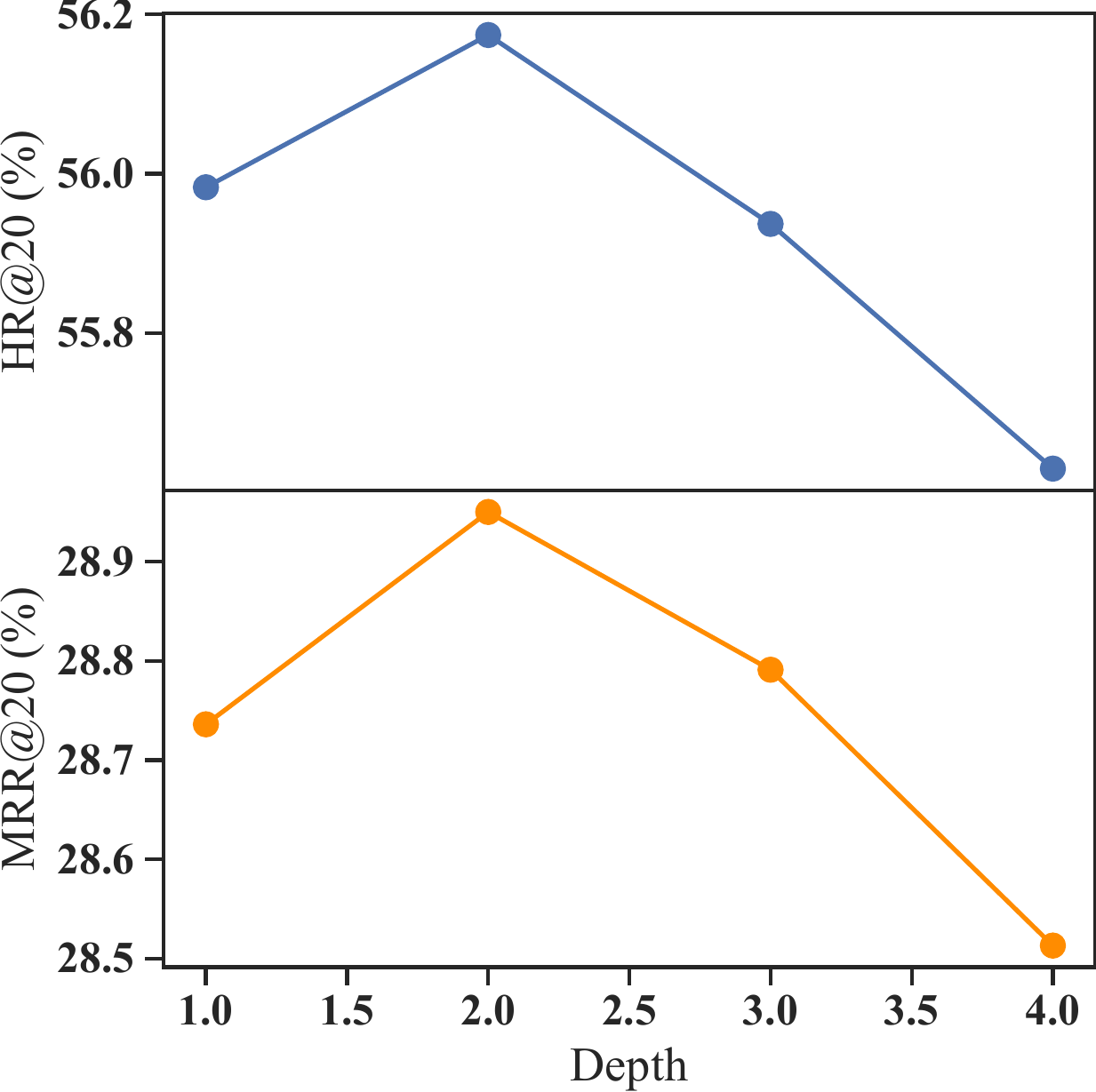}}
  \subfigure[Last.FM]{\includegraphics[width=.4\linewidth]{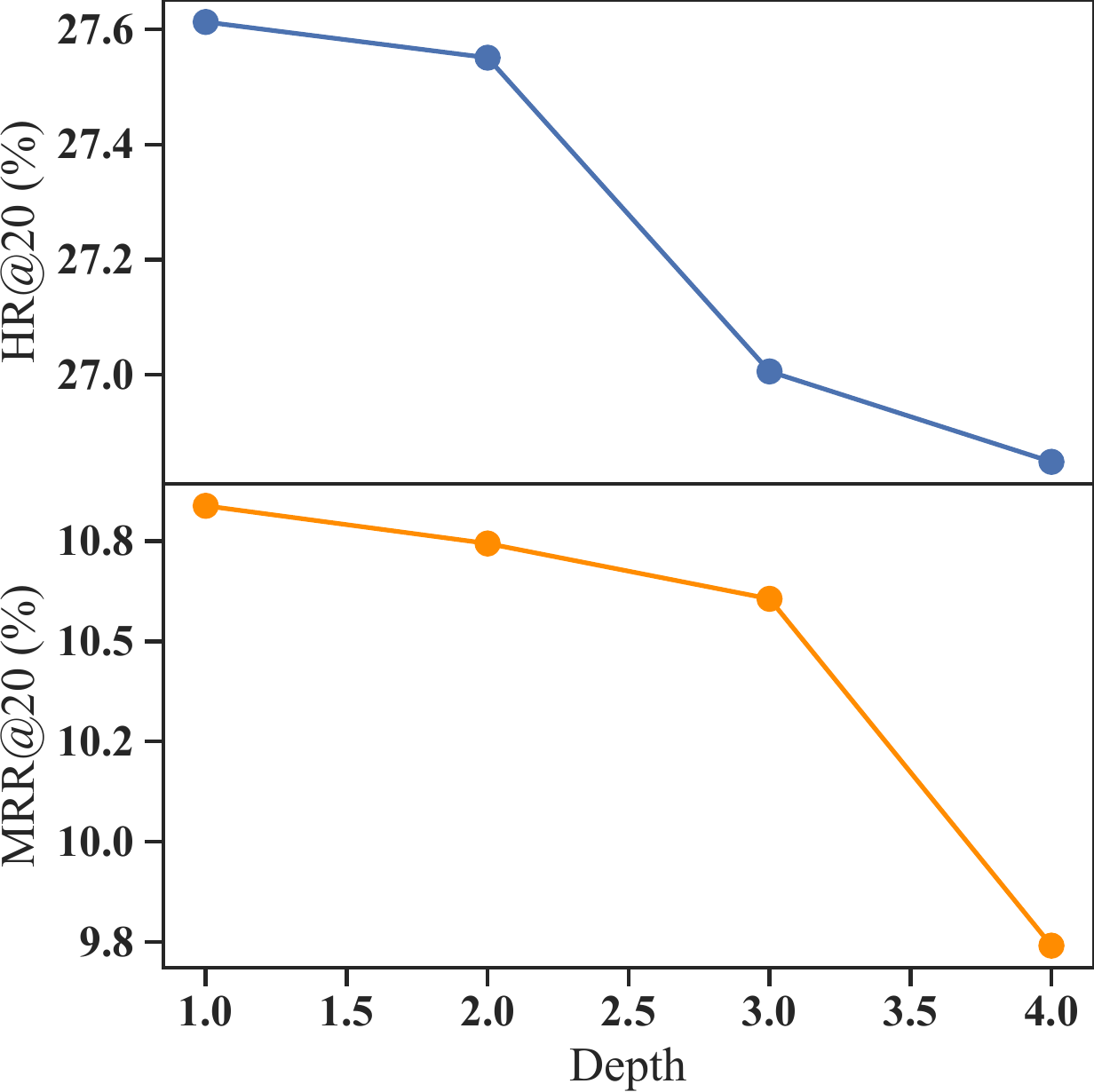}}
  \subfigure[Yoochoose1/64]{\includegraphics[width=.4\linewidth]{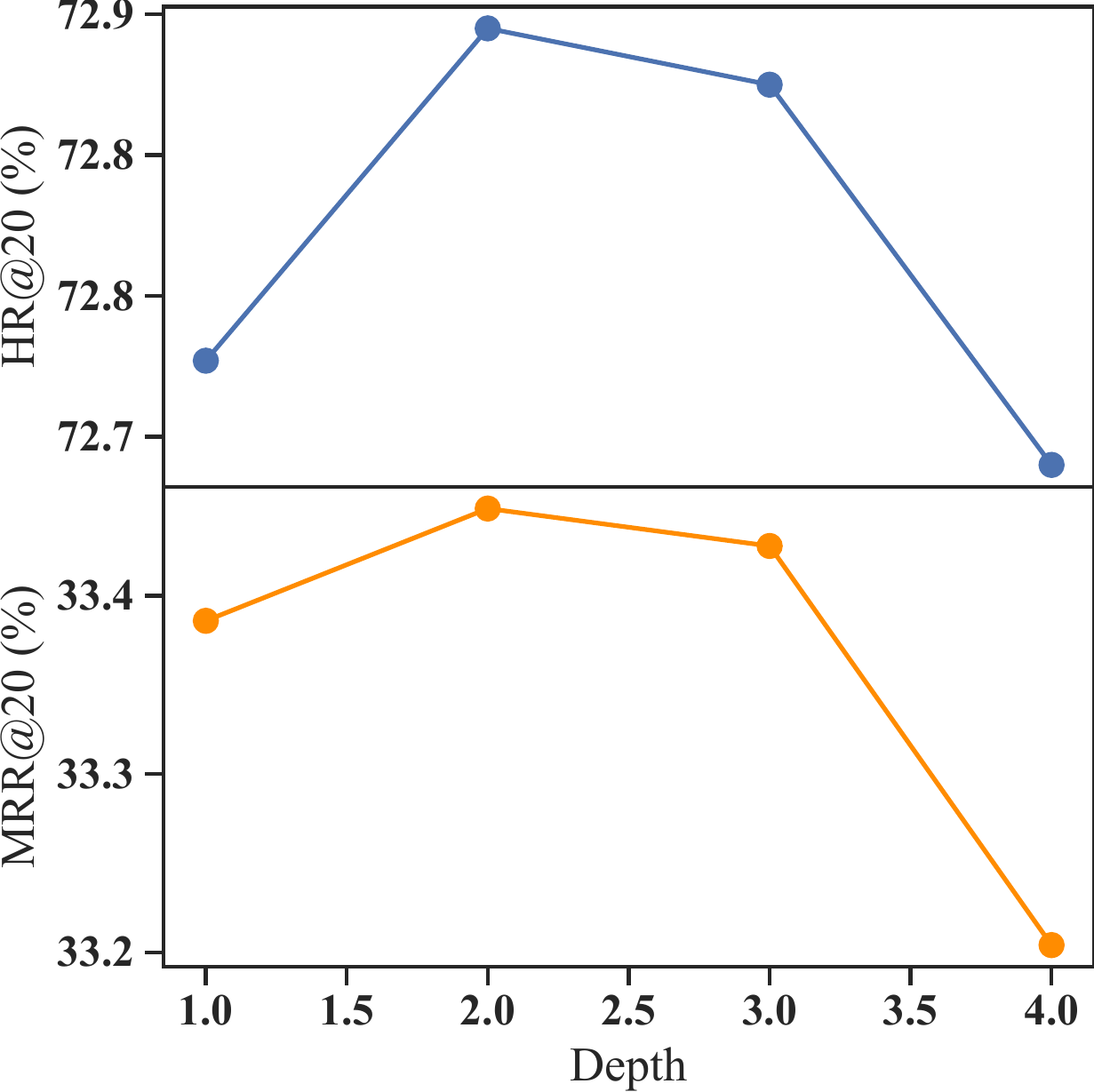}}
  \end{center}
  \caption{Results on different number of HGAT layers.} 
  \label{fig:layers}
\end{figure}

We also study the performance of different HGAT layers, and the results are shown in Figure~\ref{fig:layers}. From the figure, we find that deep models, e.g., depth > 3, suffer from over-fitting problems and achieve worse performance. Therefore, stacking more layers is not an effective method to capture long-range dependencies. However, we can alternatively extend the intent granularity level to relieve the burden of long-range dependencies.

\end{document}